\documentclass{IEEEtran}
\usepackage{cite}
\usepackage{amsmath,amssymb,amsfonts}
\usepackage{graphicx}
\usepackage{multicol}% http://ctan.org/pkg/multicols
\usepackage{textcomp,nicefrac}
\usepackage{subcaption}
\usepackage[british]{babel}
\def\BibTeX{{\rm B\kern-.05em{\sc i\kern-.025em b}\kern-.08em
T\kern-.1667em\lower.7ex\hbox{E}\kern-.125emX}}
\begin{document}
\title{Design of the Compact Processing Module for the ATLAS Tile Calorimeter}
% \title{First Prototypes of the Compact Processing Module for the ATLAS Tile Calorimeter}

\author{F. Carrió on behalf of the ATLAS Tile Calorimeter System
\thanks{This work was supported in part by the Spanish Ministry of Science and Innovation - RTI2018-094270-B-I00.}
\thanks{F.Carrió is with Instituto de Física Corpuscular (CSIC-UV), Valencia, Spain. (e-mail: fernando.carrio@cern.ch).}
\thanks{Copyright 2020 CERN for the benefit of the ATLAS Collaboration. CC-BY-4.0 license.}
}

\maketitle

\begin{abstract}
The LHC will undergo a major upgrade starting in 2025 towards the High Luminosity LHC (HL-LHC) to increase the instantaneous luminosity by a factor of 5 to 7 compared to the nominal value. %during Run-4 from 2027 to 2030 

The Phase-II Upgrade (2025--2027) will require the trigger and readout electronics of the ATLAS experiment to operate with the stringent conditions imposed by the HL-LHC. During this upgrade, both on- and off-detector readout electronics of TileCal will be completely replaced with a new data acquisition which will provide full-granularity information to the ATLAS trigger system.

%In this new readout architecture, the on-detector electronics will stream digitized data from the PMTs to 128 Compact Processing Modules operated in 32 ATCA carrier blades located in the counting rooms. The new readout system will require a total data bandwidth of 40 Tbps to read out the entire detector, where the cell energy will be reconstructed and transmitted to the ATLAS trigger system in real-time.

The Compact Processing Modules are responsible for the LHC bunch-crossing clock distribution towards the detector, configuration of the on-detector electronics, data acquisition, cell energy reconstruction, and data transmission to the TDAQ interface (TDAQi).

The CPM has been designed as an AMC form-factor board equipped with 8 Samtec FireFly modules for communication with the detector, a Xilinx Kintex UltraScale FPGA for data acquisition and processing, a Xilinx Artix 7 FPGA for slow control and monitoring, and other subsystems to generate high-quality clocks for the FPGAs and communications.%for data buffering, online digital processing and on-detector electronics control. 

The high-speed communication with the on-detector electronics is implemented via 32 GigaBit Transceiver links receiving detector data at 9.6 Gbps and transmitting commands and the LHC clock at 4.8 Gbps, while the reconstructed cell energies are transmitted to TDAQi via 4 FULL-mode links. Triggered data is transmitted through a FULL-mode link to the ATLAS TDAQ system via the FELIX network.

This paper introduces the design of the Compact Processing Modules for the ATLAS Tile Calorimeter Phase-II Upgrade and the results and experiences with the first prototypes.

%The GBTx is a radiation tolerant chip designed by CERN devoted for high-speed communication between on-detector and off-detector electronics in HEP experiments and beam lines.

\end{abstract}

\begin{IEEEkeywords}
ATLAS Tile Calorimeter(Tilecal), Data Acquisition (DAQ) systems, Field-Programmable Gate Array (FPGA), High Energy Physics, High-speed electronics.
\end{IEEEkeywords}

\section{Introduction} \label{sec:introduction}
\IEEEPARstart{T}{he} ATLAS experiment~\cite{atlas} is one of the two general purpose particle detectors at the Large Hadron Collider (LHC) at CERN. % and it has collected an integrated luminosity of 160 $fb^{-1}$ between Run-1(2011-2014) and Run-2 (2015-2018).

The Tile Calorimeter (TileCal)~\cite{tilecal} is the hadronic calorimeter of ATLAS covering the central region of the detector up to a pseudorapidity of $\eta<|1.7|$. TileCal is a sampling detector made of steel plates as absorber and scintillating tiles as active medium which provides precise measurement of hadrons, taus, jets and missing transverse energy ($E_{\mathrm{T}}^{\mathrm{miss}}$). 

TileCal is divided in three longitudinal segments along the beam axis:  one central Long Barrel (Long Barrel A and Long Barrel C) and two Extended Barrels (Extended Barrel A and Extended Barrel C). Each barrel is subdivided in 64 wedge-shaped modules for full azimuthal coverage.% each module covering $\Delta\phi = 0.1$ rads. 

\begin{figure}[h!]
\centering
\includegraphics[width=0.7\columnwidth]{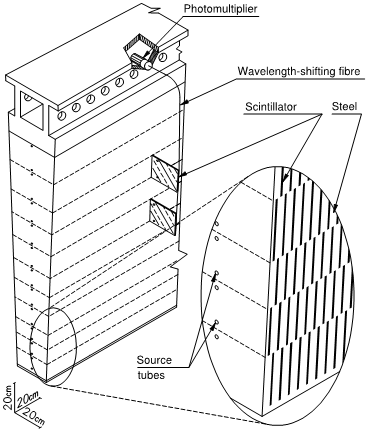}
\caption{Detail of a Tile module showing the 11 radial layers of scintillator tiles interleaved with the steel absorber, and the position of the PMTs and readout electronics~\cite{atlas}.}
\label{fig:module}
\end{figure}

The scintillating tiles are grouped into three radial layers (A,BC and D) for a total of 4672 cells in TileCal. A- and BC-cells have a dimensions of $\Delta\phi\times\Delta\eta=0.1\times0.1$, and D-cells $\Delta\phi\times\Delta\eta=0.2\times0.1$. The light generated in each cell is collected by wavelength shifting fibers at two opposite edges, and routed to two photomultiplier tubes (PMT) in the outermost part of the modules. The PMTs and the on-detector electronics are contained in extractable super-drawers, where each module hosts two drawers forming one super-drawer. Figure~\ref{fig:module} presents a drawing of a TileCal module, showing the wavelength shifting fiber routed between the scintillating tiles and the PMTs.

% Each cell is read out with two PhotoMultipliers (PMT) receiving light from opposite sides of the cell. 

%Figure \ref{fig:module} presents a detailed figure of a TileCal module showing the 11 radial layers of scintillator tiles interposed with the steel absorber, and the position of the PMTs and readout electronics.

The PMT signals are shaped and digitized at the LHC frequency (40~MHz) in the on-detector electronics with 10-bit Analog-to-Digital Converters (ADCs) and stored in pipeline memories until the reception of a Level-1 trigger acceptance signal. The triggered data is extracted from the pipeline memories, sent to derandomizer buffers and then transferred to off-detector via the Interface Card.

The Read-Out Drivers (ROD)\cite{rods} in the off-detector electronics receive the triggered data from 8 TileCal modules and transmit the reconstructed energy and time for each PMT channel to the High-Level Trigger system at a maximum average trigger rate of 100 kHz. The first stage of the off-detector electronics is composed of 32 RODs for the complete readout of the 256 TileCal modules.

In addition, copies of the analog PMT signals are grouped in towers and transmitted from the on-detector electronics to the Level-1 trigger system for particle identification and trigger decision.

%The on-detector electronics shape and digitize the signals from the PMTs every $\sim$25 ns with 10-bit Analog-to-Digital Converter (ADC) chips. Data is stored in pipeline memories until the reception of trigger acceptance signal when data is transferred to the off-detector electronics. In parallel, the PMT signals are grouped and transmitted to the ATLAS trigger system for trigger decision.

% In the back-end electronics, the main component is the Read-Out Driver (ROD)  which  performs preprocessing and gather the data coming from the front-end electronics at a maximum average trigger rate of 100 kHz. After performing the energy and time reconstruction for each channel, RODs transmit the processed data to the High Level Trigger system. A total number of 32 RODs are required for the complete read out of the Tile Calorimeter.

\section{ATLAS Phase-II Upgrade}

The High Luminosity LHC (HL-LHC) will exploit the potential of the LHC with a series of upgrades of the accelerator magnets and cavities. The HL-LHC will provide an instantaneous luminosity of $7.5\times10^{-34} \mathrm{cm}^{-2}\mathrm{s}^{-1}$ increasing the total pile-up collisions in ATLAS up to 200 simultaneous events. The HL-LHC is expected to deliver a total integrated luminosity of 4000 $\mathrm{fb}^{-1}$ in 10 years of operation.

During the ATLAS Phase-II Upgrade (2025--2027) the complete readout electronics system of TileCal will be replaced by an upgrade version using a new data acquisition architecture and trigger interfaces~\cite{TileTDR}. The new readout electronics is designed to fulfill  the radiation levels constraint under HL-LHC conditions, as well as, the demanding processing capabilities and data throughputs required by the full-digital ATLAS Trigger and Data AcQuisition (TDAQ) system.

% the readout electronics of TileCal will be completely replaced to cope with the new requirements in radiation levels of the HL-LHC and increased data bandwidths imposed by the full-digital ATLAS TDAQ system. 

%The LHC was designed to collide protons at a centre of mass energy of 14 TeV with a luminosity of $1\times10^{-34} cm^{-2}s^{-1}$ which allowed the discovery of the Higgs boson on July 2012. In order to fully exploit  its potential, the LHC plans a series of upgrades towards the High Luminosity-LHC. After these upgrades the LHC will reach an instantaneous luminosity of $7.5\times10^{-34} cm^{-2}s^{-1}$ with an expected number of collisions per bunch crossing up to 200 collisions in average and  providing a total integrated luminosity of 4000 $fb^{-1}$ in 10 years.

\begin{figure}[h!]
\includegraphics[width=1.0\columnwidth]{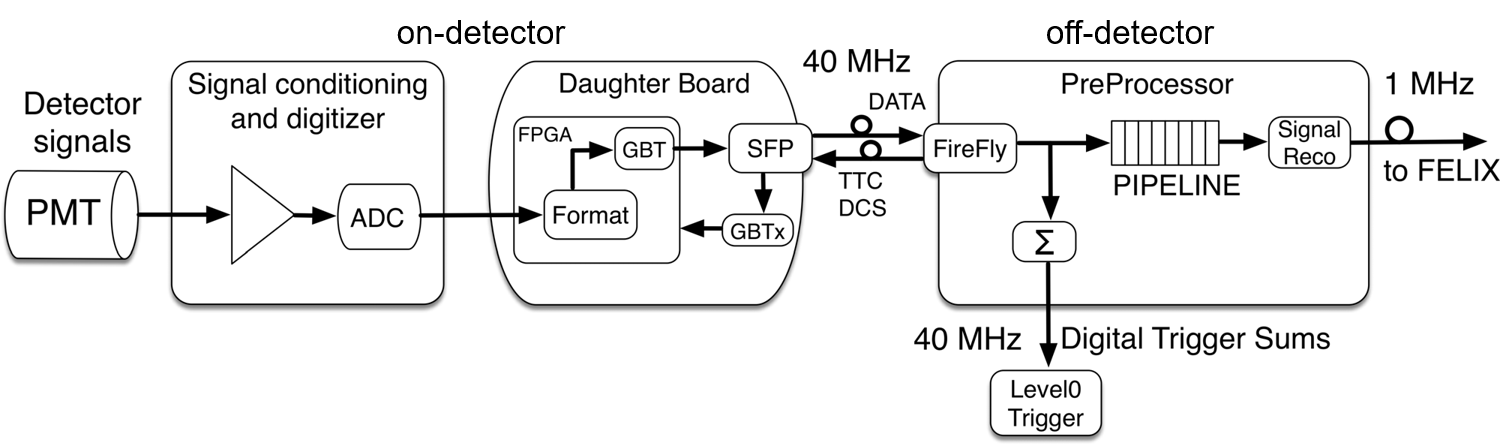}
\caption{Diagram of the Tile data acquisition architecture for the HL-LHC~\cite{TilePPr}. \label{fig:diagram}}
\end{figure}

As for the current Tile data acquisition architecture, the on-detector electronics will transmit detector data to the off-detector electronics for every bunch crossing ($\sim$25 ns). However, in the new system the data acquisition is done through 3584 optical fibers running at 9.6~Gbps requiring a total bandwidth of about 35~Tbps. Therefore, no data buffering or  processing will be done in the on-detector electronics. Figure~\ref{fig:diagram} presents a diagram of the trigger and readout architecture of TileCal at the HL-LHC.

In the off-detector electronics, 32 Tile PreProcessor~\cite{TilePPr} (TilePPr) modules will collect and process in real-time the data from 256 TileCal modules, keeping the current ratio between the number of PMT channels and RODs. The detector data will be buffered in pipeline memories capable of storing up to 10~$\mu$s of consecutive data samples, as specified in the L0-TDAQ architecture~\cite{TDAQTDR}. In parallel, the energy of each cell will be reconstructed, calibrated and transmitted in real time to the Trigger and DAQ interface module (TDAQi) which provides trigger objects to the ATLAS trigger for performing the Level-0 trigger decision with a maximum trigger rate of 1~MHz. Upon the reception of a Level-0 trigger acceptance signal the selected data is then formatted and propagated to the Front End LInk eXchange (FELIX) system~\cite{FELIX} in the ATLAS TDAQ system.

\subsection{On-detector electronics}
The upgraded TileCal modules will be segmented into identical sub-modules, called mini-drawers, to boost the availability of the readout electronics by the suppression of single-points of failure and the use of redundant power supplies. The Long Barrel modules will consist of four mini-drawers with 45 PMTs, while the Extended Barrel modules will be composed of three mini-drawers with 32 PMTs.

% Each mini-drawer forms an independent subsystem which contains all the required electronics for the operation of up to 12 PMTs, high and low voltage distribution system and high-speed optical interfaces for the communication with the off-detector electronics. 

Each mini-drawer forms an independent readout subsystem capable to operate up to 12 PMT channels. A mini-drawer is composed of a mechanical aluminum structure which supports one Mainboard, one Daughterboard, and one high voltage regulation board to read out and operate up to 12~PMT blocks equipped with FENICS front-end boards. The PMT signals are shaped and amplified in two gains by the FENICS, and digitized in the Mainboard by 12-bit dual ADCs. The digitized pulses are transferred to the Daughterboard FPGAs which create and transmit a data packet with the samples to the off-detector electronics through high-speed optical modules.

\begin{figure}[h!]
\centering
\includegraphics[width=0.7\columnwidth]{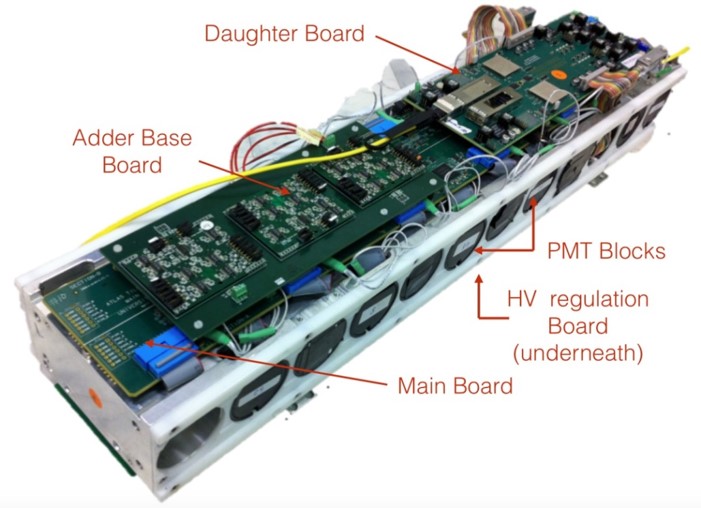}
\caption{Picture of one Tile mini-drawer containing the upgraded on-detector electronics~\cite{TileTDR}. %The red line indicates the redundancy of the electronics system.
\label{fig:md}}
\end{figure}

Figure~\ref{fig:md} presents one fully-assembled prototype of a mini-drawer populated with the upgraded on-detector readout electronics and the mechanical aluminum substructure for supporting the PMTs and the electronics.

\subsection{Off-detector electronics}

%hablamos brevemente del TilePPr y del TDAQi
The off-detector electronics of TileCal for the HL-LHC will be formed by 32 TilePPr modules and 32 TDAQi boards. 

The TilePPr module is composed of an Advanced Telecommunications Computing Architecture (ATCA)~\cite{ATCA} Carrier Blade Board (ACBB) and four FPGA-based Compact Processing Modules (CPM).

\begin{figure}[h!]
\centering
\includegraphics[width=0.8\columnwidth]{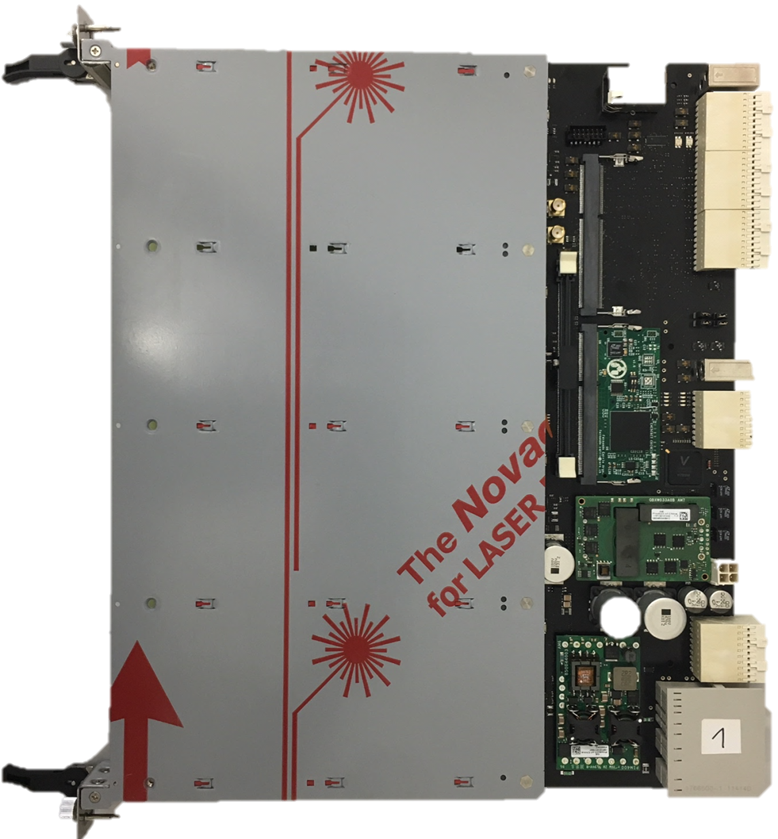}
\caption{Picture of the ATCA Carrier Base Board. \label{fig:carrier}}
\end{figure}

Figure~\ref{fig:carrier} presents a picture of the ACBB. This board has been designed as an ATCA cutaway carrier board to allow the integration of large heat sinks for the CPMs. The ACBB interconnects the CPMs with the TDAQi system through the Zone 3 connector and provides slow control capabilities through the ATCA infrastructure. The ACBB can deliver up to 400~W at 12~V through the Zone 1 connector.

In order to enhance the maintainability and upgradability of the ACBB during the entire HL-LHC lifetime, the functionalities of the ACBB are implemented into three mezzanine boards: % which can be replaced at any moment during the HL-LHC lifetime
\begin{itemize}
\item{The Tile Computer on Module (TileCoM): This board is a System-on-Chip-based board which permits remote configuration and monitoring of the on-detector and off-detector readout systems and interfaces with the Detector Control System (DCS).}
% This 204-pin SO-DIMM form factor board is equipped with a System-on-chip Xilinx ZYNQ FPGA running embedded Linux operating system. The main purpose of the TileCoM
\item{The GbE Switch: This mezzanine provides GbE communication between the CPM, TDAQi and TileCoM boards, and the ATLAS TDAQ system.}
\item{The Intelligent Platform Management Controller (IPMC): The CERN IMPC~\cite{IPMC} implements the IPMI (I for Interface) functionalities required for controlling and monitoring the cooling and power system of the ACBB.}
\end{itemize}

% As shown in the previous figure, Zone 1 connector provides up to 400~W distributing -48~V to the on-board power supplies for the generation of 12~ V, and it also interfaces the IMPC board with the SM. The Zone 2 connector is used to provide communication between the ATCA framework and the ACBB through two 10~GbE ports and two GbE ports connected to the GbE Switch board. 

% On the other hand,  while functionalities for power management, control and configuration of the boards are implemented in separated mezzanine cards. The use of separated boards allows the reduction of the complexity of the overall system and provides the ability to easily upgrade and replace them if necessary.

% The communication interfaces of the CPMs, TDAQi and the ATCA backplane, monitoring, configuration and power distribution are managed by the ACBB, while 

\section{Compact Processing Modules}

The Compact Processing Modules are the core processing and high-speed interface of the off-detector electronics of TileCal. The CPMs provide high-speed communication between the on-detector electronics and the ATLAS TDAQ system, where a total of 128 CPMs and 32 ATCA carriers will be required to read out the entire calorimeter in the HL-LHC era.

%  Detector data is buffered in pipeline memories in the FPGAs and energy and time per cell is reconstructed for every bunch crossing. 

Each CPM will operate two TileCal modules: one Long Barrel with 45 PMT channels distributed in 4 mini-drawers, and one Extended Barrel with 32 PMTs distributed in 3 mini-drawers. 

The optical high-speed interface with the on-detector electronics is implemented via four uplinks (9.6~Gbps) and two downlinks (4.8~Gbps) per mini-drawer using the GigaBit Transceiver (GBT) protocol~\cite{GBTFPGA}. The uplinks will transmit 12-bit digitized PMT signals and monitoring data to the CPMs, while the downlinks will provide Detector Control System (DCS) commands and configuration to the on-detector electronics, as well as the bunch-crossing clock for the sampling of the PMT signals. 

The energy deposited in each cell is reconstructed and calibrated in real-time from the digitized detector data. The CPM transmits the reconstructed cell energy to the TDAQi per every bunch crossing via four FULL-mode links operating at 9.6~Gbps and using 8b10b encoding~\cite{FELIX}. A maximum latency of 1.7~$\mu$s from the collision time to the reception of the computed trigger objects in the ATLAS trigger system is required to provide enough time to the trigger system for the execution of particle identification and trigger decision algorithms.

Parallel to the trigger path, the Main FPGA of the CPM stores the detector data received from the on-detector electronics and the reconstructed cell energy in RAM block-based pipeline memories, until the reception of a Level-0 trigger acceptance signal. Then, the triggered event data is transmitted to the FELIX system through high-speed links using the FULL-mode protocol for a maximum trigger rate of 1~MHz. In the opposite direction, FELIX provides to the CPMs the Trigger, Timing and Control (TTC) information and bunch-crossing clock via a GBT link at 4.8~Gbps. All high-speed interfaces operate with fixed and deterministic latency. %to ensure

\subsection{Hardware design and connectivity}
% \subsubsection{FPGAs}

% Kintex UltraScale+FireFly

\begin{figure}[h]
\centering
    \includegraphics[width=1.\columnwidth]{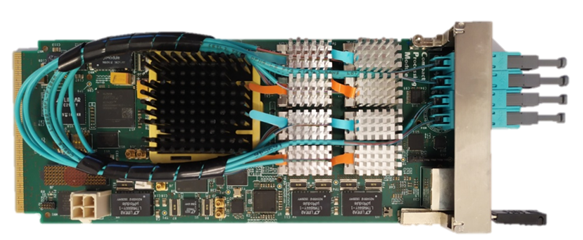}
    \caption{Picture of the first prototype of the Compact Processing Module.}
    \label{fig:CPM}
\end{figure}

The CPM has been designed as single full-size Advanced Mezzanine Card (AMC) hosting one Xilinx UltraScale FPGA (Main FPGA), one Xilinx Artix FPGA (Control FPGA) and eight Samtec FireFly optical modules. The prototypes of the CPM assembled with the 8 Samtec FireFly modules is presented in Figure~\ref{fig:CPM}.

\begin{figure*}[h!]
\centering
    \includegraphics[width=1.25\columnwidth]{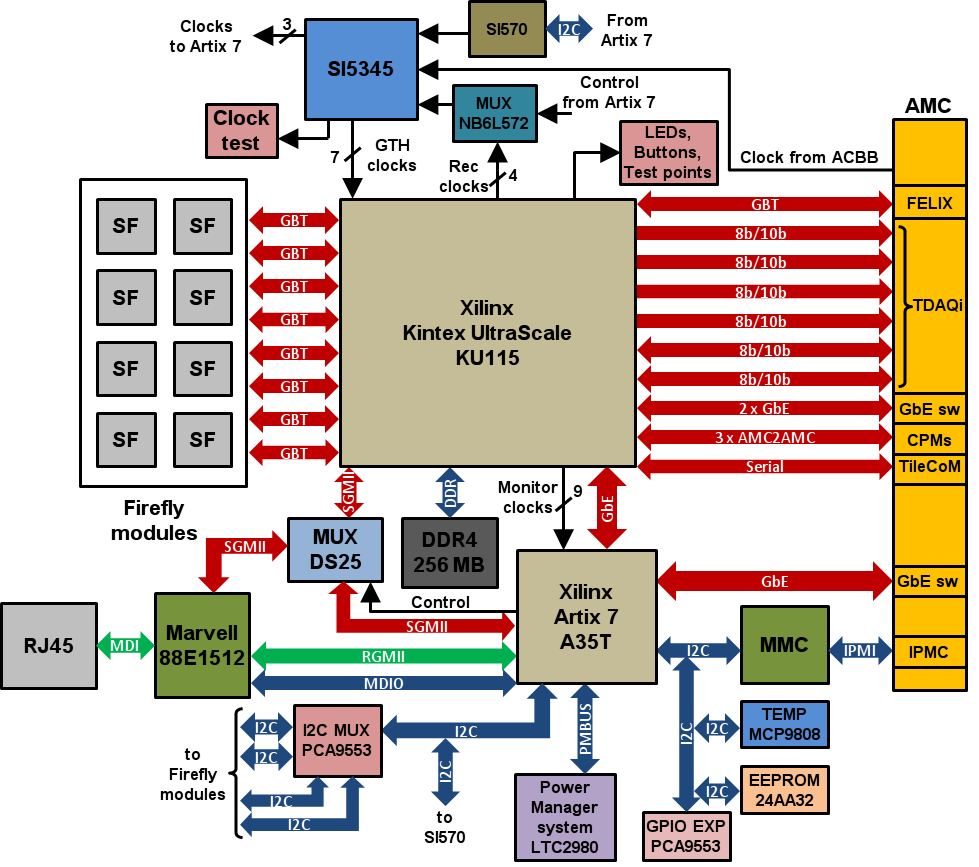}
    \caption{Block diagram of the Compact Processing Module.}
    \label{fig:bdCPM}
\end{figure*}

Figure~\ref{fig:bdCPM} shows a complete block diagram of the CPM and its interconnections. The core processing of the CPM is a Kintex UltraScale FPGA, called Main FPGA. The first CPM prototypes mount a Kintex KU085-1A1517C FPGAs with 48 GTH MultiGigabit Transceivers (MGT) capable of operating up to 12.5~Gbps data rates. The final version of the CPM to be installed in ATLAS will be equipped with a Kintex KU115-1A1517C FPGAs in order to provide a large margin of resources for the implementation of more complex energy reconstruction algorithms.

The Main FPGA provides a bidirectional optical communication path with the on-detector electronics via 32 MGTs connected to 8 FireFly modules, while 5 MGTs interface with the TDAQi and FELIX systems through the AMC connector. The remaining MGTs are used for the communication with the GbE Switch and to the other AMCs in the ACBB. 

The Modular Management Controller (MMC) of the CPM also communicates through the AMC connector with the IPMC in the carrier to manage the hot swap power sequence and for transmitting health monitoring data. 

%The ACBB can deliver up to 60 W at 12 V through the AMC connector.

% A Modular Management Controller (MMC) board provides the operation of the basic services to the prototype when connected to the ATCA system using the Intelligent Platform Management Interface (IPMI) protocol. The MMC manages the power connection of the TilePPr with the carrier and monitors the temperature sensors on the board.
%Dominant factors in this choice are the number of GTH transceivers, and the amount of logic and DSP resources.

% The rest of the GTH transceivers are used for the GbE communication with the ACBB mezzanine boards, communication between Kintex UltraScale and Artix 7, and for test purposes. The test connectors, DDR4 memory interface, and clock forwarding to A7 for monitoring are connected to GPIO pins.

% Artix 7 FPGA
The slow control and monitoring of the peripherals and power modules relies on a Xilinx XC7A35T-1CSG325C, called Control FPGA. The Control FPGA receives the detector monitoring data from the Main FPGA and transmit it to the ATLAS DCS system through the TileCoM board. Furthermore, the Control FPGA is intended to implement a phase monitoring circuit based on the Digital Dual Mixer Time Difference phase detector~\cite{OSUS} for the monitoring of phase variations in the distributed clock with a precision of 30~$\text{ps}_{\text{RMS}}$.

% MMC and I2C, sensors, etc

% 

%JTAG chains

% AMC and interconnectivity

% Each CPM is connected to three GbE ports of the GbE Switch board and two JTAG chains controlled by the TileCoM. The first JTAG is devoted for the remote or local configuration of the CPM FPGA, while the second one is used for remote programming the Daughter Board FPGAs over the fiber optics. Finally, the Zone 3 connector provides up to 32 point-to-point connections to interface the CPMs and the TDAQi for the communication with the FELIX system and to the FPGAs in charge of the trigger processing.

% The CPM receives 12 V from the ACBB through the AMC connector. According to the ATCA standard specifications [4], the ACBB can deliver a maximum power of 80 W per single AMC board. An auxiliary 3.3 V is also provided to power the MMC board during the initial power negotiation between ACBB and the CPM.

\subsection{Clocking circuitry}

The CPM is responsible for the distribution to the on-detector electronics of the recovered bunch-crossing clock from the FELIX input stream. The designed clock architecture in the CPM permits to extract the recovered clock directly from the MGTs using the dedicated output clock buffers OBUFDS\_GTE3 and to clean it with a Si5345 jitter cleaner before routing it back onto the Main FPGA through the transceiver reference clock pins. This clock routing introduces lower levels of jitter than buffering the clock through regular general purpose input/ouput (GPIO) pins to the jitter cleaner~\cite{Mendes}.

The jitter cleaner outputs six copies of the clean version of the bunch-crossing clock to the MGT banks of the Main FPGA, so that all MGTs can operate using the same reference clock. Additional copies of the bunch-crossing clock are transmitted to the Control FPGA for phase monitoring purposes, and to MMCX connectors to measure the quality of the clock.

% for the digitization of the PMT signals
% Figure~\ref{fig:clock} presents a complete block diagram of the clock circuitry of the CPM which describes interconnections and clocking options.

% \begin{figure}[h]
% \centering
%     \includegraphics[width=1.\columnwidth]{figs/clock.png}
%     \caption{Picture of the first Compact Processing module prototype.}
%     \label{fig:CPM}
% \end{figure}

%Lastly

% The cleaned copies of the recovered clock are driven to the GTH banks of the Kintex UltraScale, Artix 7, and to a pair of on-board MMCX connectors for testing and validation purposes. All GTH banks of the KU can operate using the recovered clock. 
The clocking circuitry can be configured to provide three possible clock inputs to the Si5345 jitter cleaner: %scenarios
\begin{itemize}

\item Local programmable oscillator Si570: Used for the initialization of the transceivers and lab tests.
\item Recovered clock from FELIX: This clock is buffered through an On Semiconductor NB6LQ572 multiplexer clock. The clock multiplexer permits to choose between different copies of the clock recovered with the MGTs.
\item Clock from the ACBB: This clock is intended for lab tests and qualification purposes during the final production of the CPMs.
\end{itemize}

% Note that the fourth clock input (IN3) of the SI5345 jitter attenuator is connected to the tenth output (OUT9) to enable the 0-delay mode.
% The A7 FPGA carries the control and programming of the jitter cleaner, as well as the control of the multiplexer clock. The clock distribution system is equipped with two on-board 100 MHz oscillators to provide the FPGA’s banks with reference clocks.

%Dibujo del clock distribution

% \begin{multicols}{2}
%   \lipsum[1-2]
% \end{multicols}
% \begin{figure*}[h]
% \centering
%   \includegraphics[width=1.25\columnwidth]{figs/blockdiagram.png}
%   \caption{This is a tiger.}
% \end{figure*}
% \begin{multicols}{2}
%   \lipsum[3-4]
% \end{multicols}

\subsection{PCB design}

% The PCB  design has been carried out with the careful selection of dielectric materials combined with signal and power integrity simulations. 

The PCB comprises 14 layers, where 6 are devoted to power distribution and 8 layers are dedicated for high-speed and control signals. The total PCB thickness is 1.6 mm $\pm$10\%, compliant with the AMC standard. Isola FR408HR was selected as dielectric material due its cost-effective, low dielectric constant ($\epsilon_r\simeq$ 3.68) and low loss tangent ($\text{tan}\delta\simeq$ 0.01) compared with other standard FR4 materials. 

%The PCB interconnects were designed to match 100~$\Omega$ for differential signals and 50~$\Omega$ for single-ended signals. 

%which reduces the signal losses at high frequency when
%The single AMC CPM hosts eight Samtec FireFly modules with 4 high-speed links each for the communication with the DaughterBoards in the on-detector electronics; a Xilinx Kintex UltraScale FPGA data buffering, digital processing and control; an Ethernet port; and interconnections to communicate with the TDAQi through the ATCA carrier connectors.

% \begin{figure}[h!]
% \centering
% \includegraphics[width=1\columnwidth]{figures/cpm.png}
% \caption{Layout of the Compact Processing Modules showing the internal high-speed layers. \label{fig:cpm}}
% \end{figure}

During the PCB design, signal and power simulations were performed with ANSYS Electromagnetics Suite to verify the signal integrity of the high-speed lines and power planes. Several studies were carried to minimize the insertion loss and impedance discontinuities along the high-speed paths introduced by differential vias and DC-blocking capacitors.

%
% to detect and mitigate any problems degrading the signal quality.
%Finally, the high-speed lines of the final layout were simulated using the Xilinx IBIS-AMI models to reproduce a best real-close scenario and optimize the links.

\begin{figure}[h!]
\centering
   \includegraphics[width=0.9\columnwidth]{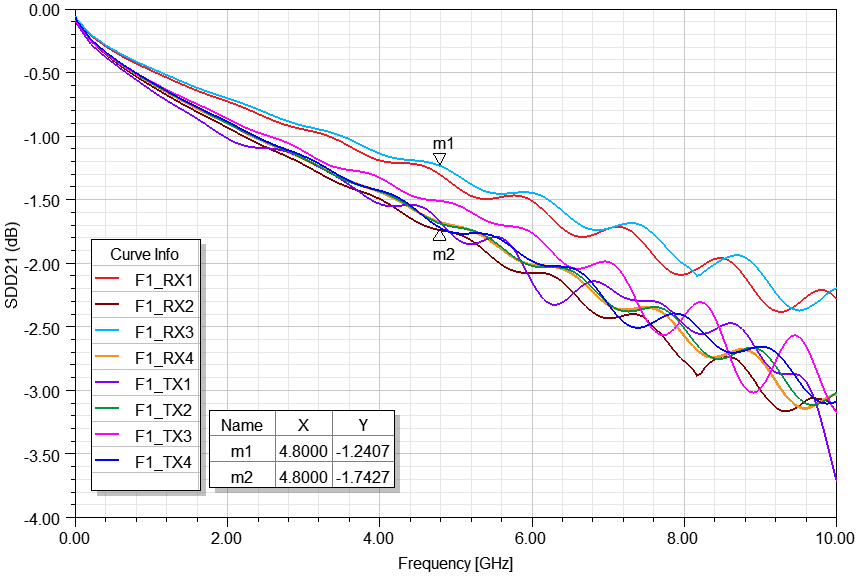}

\caption{Simulated differential insertion loss for the transmission and reception interconnects between the FPGA and one of the Samtec FireFly modules.}
\label{fig:sdd21}
\end{figure}

Figure \ref{fig:sdd21} presents the simulation results of the differential insertion loss (SDD21) for the interconnects of the Samtec FireFly module farthest from the Main FPGA. The simulated high-speed interconnects have a maximum length of 7.65~cm.

% and were designed to operate at a minimum data rate of 9.6~Gbps

As can be observed in the results, the insertion loss at the Nyquist frequency (4.8 GHz) is between $-1.24$~dB and $-1.74$~dB, ensuring signal integrity on the high-speed interconnects between the FireFly modules and the Main FPGA. %The simulated channel attenuations are compliant with the recommendations described in the SFF-8431 standard~\cite{SFF} for 10~Gbps interconnections.

\section{Test and Verification} %or prototype test results

This section presents the test results obtained during the verification of the first CPM prototypes. These tests included the qualification of the high-speed links to the on-detector electronics and the clock distribution capabilities of the CPM.

\subsection{Link qualification tests}\label{sec:BER}

The 32 optical links connected to the Main FPGA have been qualified with the Xilinx IBERT IP core~\cite{IBERT} through Bit Error Rate (BER) tests running the links at 9.6~Gbps.

%The incoming data throughput to the CPM could reach up to 307.2~Gbps (32 channels at 9.6~Gbps). 

The setup consisted in two separated CPMs transmitting and receiving $2^{31}-1$ Pseudo-Random Bit Sequence (PRBS31) data patterns during seven days. No errors were found in any link providing a BER better than $5\cdot10^{-16}$ per link with a Confidence Level (CL) of 95~\%, and a total combined BER better than $1.6\cdot10^{-17}$ with a CL of 95~\%. 

Furthermore, the power consumption during the test was about 32~W. The CPM consumption during operation is expected to increase by 12~W due to additional firmware. This estimation was obtained from the Xilinx Power Estimator tool and the expected resource use for the final firmware. These preliminary estimations predict that the power consumption of the CPM will be within the power consumption budget of 60~W per AMC specified by the ATCA standard~\cite{ATCA}.

% Consequently, the total power consumption of the CPM during the operation is expected to be lower than 40~W, which represents a margin on power budget of 33\%.
%power budget, margin

\subsection{Signal integrity validation}

As part of the validation tests of the CPMs, the eye diagrams were measured at the output of the Samtec FireFly modules using a DCA-X~86100D sampling oscilloscope from Keysight. The measurements were done transmitting a PRBS31 data pattern at data rates of 4.8~Gbps and 9.6~Gbps.

\begin{figure}[h]
\centering
\begin{subfigure}[b]{0.85\columnwidth}
  \centering
  \includegraphics[width=0.85\columnwidth]{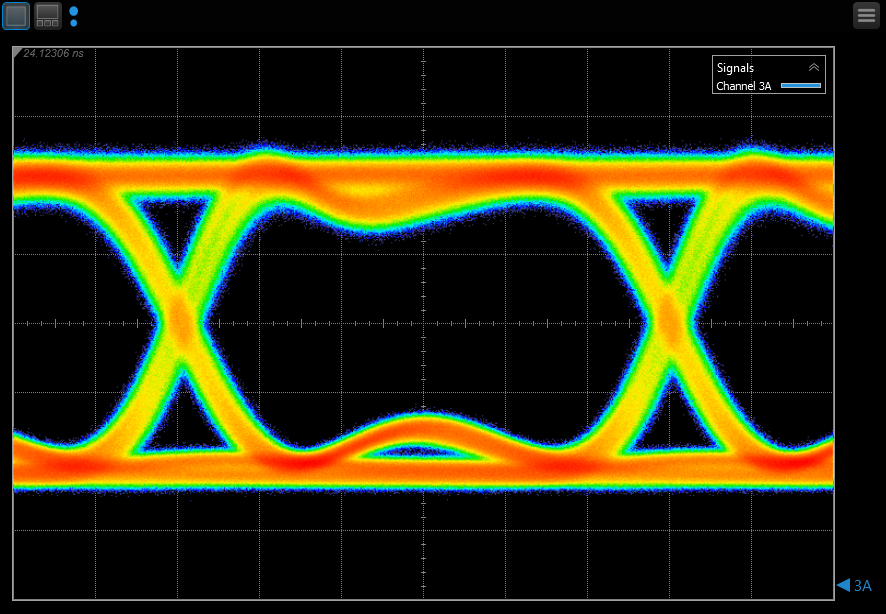}
  \caption{}
  \label{fig:Ng1} 
\end{subfigure}

\begin{subfigure}[b]{0.85\columnwidth}
  \centering
  \includegraphics[width=0.85\columnwidth]{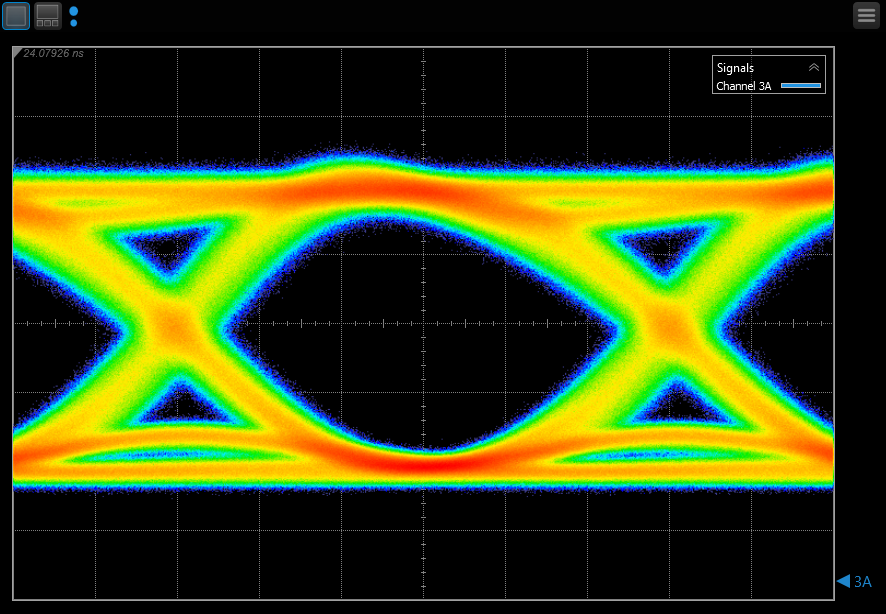}
  \caption{}
  \label{fig:Ng2}
\end{subfigure}
\caption{Measured output eye diagrams for 4.8~Gbps (a) and 9.6~Gbps (b) data rates.}
\label{fig:eye}
\end{figure}

Figure~\ref{fig:eye} shows the results of the eye diagram measurements for 4.8~Gbps (a) and for 9.6~Gbps (b). In both cases, the eye diagrams are widely opened without signs of excessive jitter, presenting satisfactory noise margins.
%noise margins
%jeopardize?
%A block diagram of the setup for these measurements is presented in Figure~\ref{fig:setup2}

Signal quality parameters were extracted from the eye diagram data as part of the validation tests. Table \ref{tab:eye} summarizes the extracted jitter values, eye height and width.
%, where each parameter corresponds to mean value of 1000 measurements.

% \usepackage{float}
% \restylefloat{table}

\begin{table}[h]
\centering
\label{tab:jitter}
\begin{tabular}{ccc}
% \textbf{}          & \multicolumn{2}{c}{\textbf{Iterations}}                                   \\ \cline{1-3} 
\hline
\textbf{Parameter} & \textbf{4.8~Gbps}            & \textbf{9.6~Gbps}\\ \hline
\hline
Eye Width              & 191.84 ps  & 85.92 ps\\
Eye Height             & 73.15\%    & 87.10\%\\
RJ (rms)               & 1.94 ps    & 1.84 ps\\
DJ ($\delta$-$\delta$) & 2.40 ps    & 5.85 ps\\
TJ ($10^{-12}$)        & 28.84 ps   & 30.96 ps\\
TJ ($10^{-13}$)        & 30.09 ps   & 32.15 ps\\
TJ ($10^{-14}$)        & 31.30 ps   & 33.28 ps\\
TJ ($10^{-15}$)        & 32.45 ps   & 34.38 ps\\
TJ ($10^{-16}$)        & 33.56 ps   & 34.43 ps\\

% Eye Width              & 191.84 ps $\pm$ 297 fs   & 85.92 ps $\pm$ 352 fs\\
% Eye Height             & 73.15\% $\pm$ 0.51\%     & 87.10\% $\pm$ 0.19\% \\
% RJ (rms)               & 1.94 ps $\pm$ 319 fs     & 1.84 ps $\pm$ 177 fs\\
% DJ ($\delta$-$\delta$) & 2.40 ps $\pm$ 1.80 ps    & 5.85 ps $\pm$ 1.19 ps\\
% TJ ($10^{-12}$)        & 28.84 ps $\pm$ 2.67 ps   & 30.96 ps $\pm$ 1.37 ps\\
% TJ ($10^{-13}$)        & 30.09 ps $\pm$ 2.86 ps   & 32.15 ps $\pm$ 1.48 ps\\
% TJ ($10^{-14}$)        & 31.30 ps $\pm$ 3.05 ps   & 33.28 ps $\pm$ 1.59 ps\\

\end{tabular}
\caption{Signal quality parameters extracted from the measured output eye diagrams at 4.8~Gbps and 9.6~Gbps.}
\label{tab:eye}
\end{table}

% \begin{table}[h]
% \centering
% \label{tab:jitter}
% \begin{tabular}{ccc}
% % \textbf{}          & \multicolumn{2}{c}{\textbf{Iterations}}                                   \\ \cline{1-3} 
% \hline
% \textbf{Parameter} & \textbf{4.8~Gbps}            & \textbf{9.6~Gbps}\\ \hline
% \hline
% Eye Width          & 181.70 ps $\pm$ 279 fs                 & 78.60 ps $\pm$ 127 fs \\
% Eye Height        & 59.3\% $\pm$ 0.1\%                 & 81.3\% $\pm$ 0.04\% \\
% Jitter[RMS]        & 4.15 ps $\pm$ 29.9 fs                 & 4.31 ps $\pm$ 18.4 fs \\
% Jitter[p-p]        & 24.92 ps $\pm$ 1.6 ps                 & 31.54 ps $\pm$ 1.6 ps \\

% \end{tabular}
% \caption{Integrated RMS jitter extracted from the measured phase noise plots.}
% \label{tab:eye}
% \end{table}

% There is some overshoot and undershoot which reduces the opening ratio, although this symptom does not represent a risk for this application, it can be minimized by a fine configuration of the GTH transmitter features. As observed in Figure, this overshoot and undershoot are not present at 9.6 Gbps due to the losses .

The signal quality at the input of the CPM was also studied using the built-in capabilities of the Xilinx transceivers. For these measurements, the same setup presented in section~\ref{sec:BER} is used. The receiver of the MGTs includes a second independent sampler after the equalizer which runs in parallel with the Clock and Data Recovery (CDR) data sampler. The results from both samplers are compared to perform the BER tests and to compose eye diagrams.

\begin{figure}[h!]
\centering
   \includegraphics[width=0.9\columnwidth]{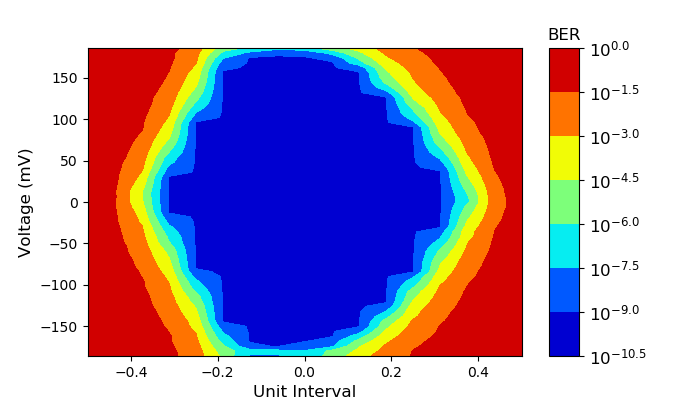}

\caption{Eye diagram measured at the input of the Xilinx GTH MGT for a BER better than $10^{-9}$ per sampling point.}
\label{fig:eye2}
\end{figure}

Figure~\ref{fig:eye2} presents the eye diagram at the input of the FPGA for a BER better than $10^{-9}$ per sampling point and a data rate of 9.6~Gbps. The eye diagram is scanned across vertically and horizontally by controlling the clock phase of the sampler  and the offset voltage.
%in steps of 1.625~ps
%in steps of 1.5~mV
% Figure~\ref{fig:eye2} presents the bathtub curve, a eye margin of 0.62 UI for a BER better than $1\cdot10^{-12}$ per point measured. 

An horizontal BER test scan was performed by shifting the phase clock of the second sampler horizontally across the eye, and setting the voltage offset to the center of the eye.
% in steps of 1.5~mV

\begin{figure}[h!]
\centering
   \includegraphics[width=0.9\columnwidth]{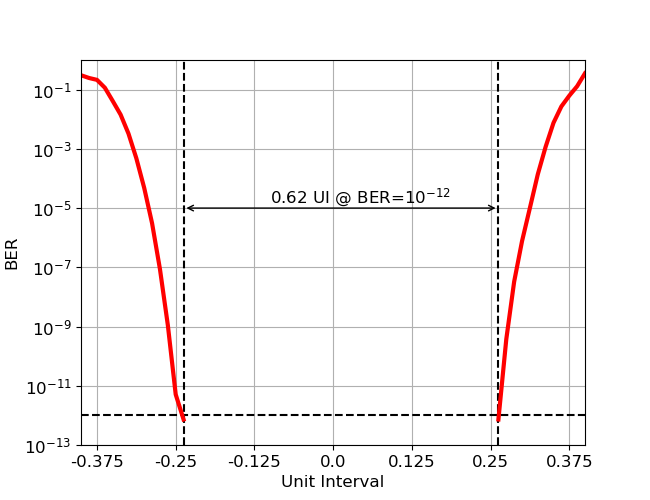}
\caption{Bathtub curve corresponding to an horizontal BER test scan up to $10^{-12}$ per sampling point.}
\label{fig:bath}
\end{figure}

The resulting BER curve is a bathtub curve, so-called because of its characteristic shape, shown in Figure~\ref{fig:bath}. The BER is below $10^{-12}$ at the center of the eye and increases when getting closer to the eye crossings.

The distance between the left and the right curves defines a noise margin of 0.62~Unit Interval ($\sim$64.58~ps) at the specified BER level of $10^{-12}$. All channels were verified and validated showing similar quantities.

% The resulting BER curve is shownschematically in Fig. B.9(a). The BER is low when sampling at the center of theeye and goes up when approaching the eye crossings to the left and right. Thiscurve is known as thebathtub curvebecause of its characteristic shape.

% hehorizontal eye margin is the separation of the two sides of the bathtub curve ata specified BER level (see Fig. B.9(a)). For example, in the 10-GbE standard, the horizontal eye margin is specified for a BER of 10−12

% \begin{figure}[h!]
% \centering
% \begin{subfigure}[b]{1\columnwidth}
%   \centering
%   \hspace{10mm}%
%   \includegraphics[width=0.9\columnwidth]{figs/eye.png}
%   \caption{}
%   \label{fig:Ng1} 
% \end{subfigure}
% \begin{subfigure}[b]{1\columnwidth}
%   \centering
%   \includegraphics[width=0.7\columnwidth]{figs/bath.png}
%   \caption{}
%   \label{fig:Ng2}
% \end{subfigure}
% \caption{(a) Eye diagram measured at the input of the Xilinx GTH transceiver for a BER better than $10^{-9}$ per measurement; and (b) the corresponding Bathtub curve resulting from a horizontal BER test scan up to $10^{-12}$ per measurement.}
% \label{fig:eye2}
% \end{figure}

\subsection{Clock qualification tests}

\begin{figure}[h]
\centering
    \includegraphics[angle=0,origin=c, width=1.0\columnwidth]{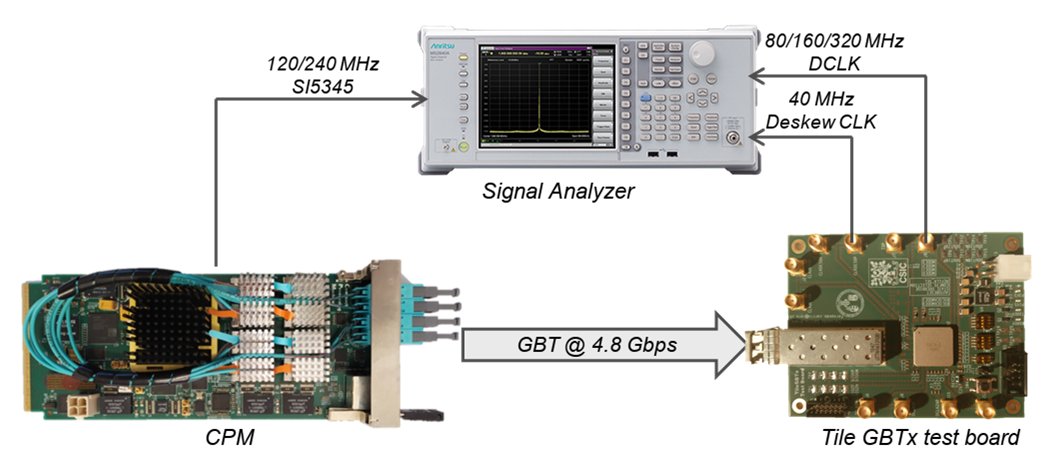}
    \caption{Diagram of the test bench for the clock qualification tests.}
    \label{fig:tilegbtx}
\end{figure}

In order to study the performance of the clock distribution system, phase noise measurements were carried with a MS2840A signal analyzer from Anritsu at room temperature. Figure~\ref{fig:tilegbtx} presents a block diagram of the setup for the clock qualification tests. The setup consisted in a CPM with a GBT link transmitting at 4.8~Gbps to a GBTx chip~\cite{GBTx} in a test board via a SFP module. 
% The CPM recovered the bunch-crossing clock was recovered from a mini-FELIX card (Xilinx VC709 evaluation board)~\cite{miniFELIX} via a GBT link at 4.8 Gbps.

%% The CPM was synchronized to the bunch-crossing clock, by recovering it from a mini-FELIX card (Xilinx VC709 evaluation board)~\cite{miniFELIX} via a GBT link at 4.8 Gbps.

% The CPM was interfaced with a mini-FELIX card (Xilinx VC709 evaluation board)~\cite{miniFELIX} via a GBT link at 4.8 Gbps to recover the bunch-crossing clock from the input data stream. 

%n and several SMA connectors and test points

The GBTx chip is a 130~nm CMOS rad-hard device intended for high-speed communications and slow control in High Energy Physics experiments. This chip is able to recover the bunch-crossing clock from the input serial link providing up to 8 independent output clocks with programmable phase and up to 40 clocks with a fixed phase. Both types of clocks can be configured to buffer 40, 80, 160 and 320~MHz clocks. 

However, the phase-programmable clocks present higher levels of random jitter than the e-link clocks, showing a phase noise peak around 1 MHz introduced by the internal delay lines used for shifting the phase. For this reason, the e-link clock is preferred over the Phase-Shift clock (PS clock) as reference clock for driving the MGTs of the Daughterboard FPGAs. Despite its higher jitter, the PS clock is used as input clock to the Mainboard ADCs since the phase-tuning capability is desirable to minimize the error of the real time energy reconstruction algorithms and for pulse characterization purposes and calibration.
%The Versatile Link Demo Board (VLDB)
%https://inspirehep.net/files/1111dfd050f49fdfef64d6be023f0242

%showing 
%Sería interesante poner un diagrama de ojo de cómo se recibe la DB.
% explicar test bench

\begin{figure}[h]
\centering
    \includegraphics[width=0.9\columnwidth]{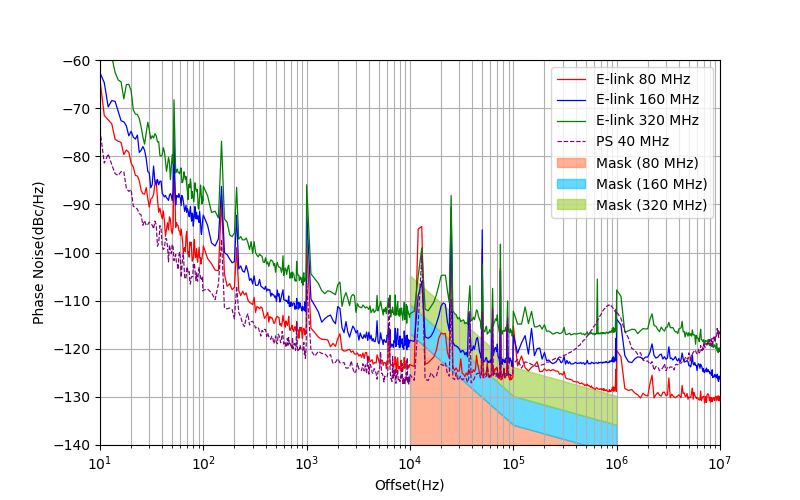}
    \caption{Phase noise plot of the e-link and PS clocks under study.}
    \label{fig:pn_all}
\end{figure}

Figure \ref{fig:pn_all} presents the measured phase noise plot of the e-link and PS clocks for different frequencies, as well as the corresponding phase noise mask provided by the FPGA manufacturer to ensure minimal jitter in the MGT outputs~\cite{KUdatasheet}. In this test, the clock source was the on-board Si5345 jitter cleaner driven by the on-board Si570 oscillator.

% In this test, the clock recovered from mini-FELIX was driven from the Main FPGA to the on-board Si5345 jitter cleaner using an OBUFDS\_GTE4 buffer.

%or for jitter reduction

%For this study, the GBTx chip was configured in transceiver mode and it was connected to the CPM through a GBT link injecting data at 4.8 Gbps.

The e-links were configured to provide recovered clocks with frequencies of 80~MHz, 160~MHz and 320~MHz. The phase noise of the 40~MHz e-link clock is not shown for simplicity, since 40~MHz is outside the frequency range for the reference clock of the MGTs. Similarly, only the phase noise of the 40~MHz PS clock is presented because it is the only possible input clock frequency for the 12-bit ADCs on the Mainboard.

As observed in Figure~\ref{fig:pn_all}, the phase noise values of all the e-link clocks exceeds the phase noise limits for the reference clocks addressed by the FPGA manufacturer. However, stable communication between the CPM and the Daughterboard~v5~\cite{Eduardin} is achieved using the GBT protocol at 4.8~Gbps for the downlink and 9.6~Gbps for the uplink.

%though the measured phase noise exceeds the Xilinx recommendations for driving the GTH transceivers

% \begin{figure}[h]
% \centering
%     \includegraphics[width=0.9\columnwidth]{figs/deskew240.png}
%     \caption{Phase noise xxxx.}
% \end{figure}

Table~\ref{tab:jitter2} shows the integrated RMS jitter extracted from the phase noise measurements of the two types GBTx output clocks and for all frequencies. The RMS jitter was obtained integrating the phase noise response over the entire bandwidth (from 10~Hz to 10~MHz) and over the sensitive bandwidth of the MGT (from 10~kHz to 1~MHz). %The spurs present in the data are due to the power supply switching noise

\begin{table}[h!]
\centering
\begin{tabular}{ccc}
% \textbf{}          & \multicolumn{2}{c}{\textbf{Iterations}}                                   \\ \cline{1-3} 
\textbf{Reference}            & \textbf{10~Hz--10~MHz}  & \textbf{10 kHz--1 MHz}\\ \hline
\hline
E-link 40 MHz                        & 6.43 ps & 3.60 ps\\
E-link 80 MHz                        & 4.57 ps & 2.91 ps\\
E-link 160 MHz                      & 3.96 ps & 1.75 ps\\
E-link 320 MHz                       & 4.02 ps & 1.43 ps\\
PS 40 MHz                         & 20.10 ps & 9.88 ps\\
PS 80 MHz                         & 12.81 ps & 9.78 ps\\
PS 160 MHz                         & 12.21 ps & 9.57 ps\\
PS 320 MHz                         & 12.23 ps & 9.66 ps\\
\end{tabular}
\caption{Integrated RMS jitter extracted from the measured phase noise plots at the outputs of the GBTx.}
\label{tab:jitter2}
\end{table}

The RMS jitter of the 80/160/320 MHz PS clocks for the entire bandwidth is about three times higher than in the e-link clocks. In the frequency range specified by the manufacturer, the RMS jitter is up to seven times higher in the PS clocks.

Finally, the phase noise at the outputs of the Si5345 jitter cleaner of the CPM was measured. Two different clock frequencies of 120 MHz and 240 MHz were considered as reference clocks for the implementation of the GBT protocol. 
%generated by the local Si570 clock oscillator 

% Finally, the phase noise at the outputs of the Si5345 jitter cleaner of the CPM was measured. Two different clock frequencies of 120 MHz and 240 MHz were considered as reference clocks for the implementation of the GBT protocol. 

\begin{figure}[h!]
\centering
    \includegraphics[width=0.9\columnwidth]{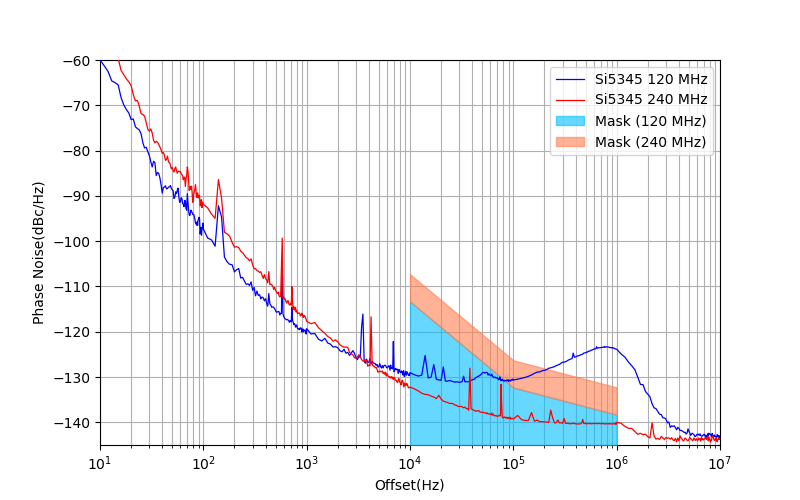} %original.png
    \caption{Phase noise plot of the 120 MHz and 240 MHz at the output of the Si5345 jitter cleaner.}
    % \caption{Phase noise plot of the local reference clocks and  at the output of the Si5345 jitter cleaner.}
    \label{fig:ph_orig}
\end{figure}

Figure~\ref{fig:ph_orig} shows the phase noise measurements for the two reference clocks in the CPM and the corresponding phase noise masks. 

%Figure~\ref{fig:ph_orig} shows the phase noise measurements at the output of the Si5345 for the two possible reference clocks, the 240~MHz recovered clocks from mini-FELIX for the GPIO and OBUFDS\_GTE4 options, and the corresponding phase noise masks.

%, the recovered clocks from mini-FELIX driven to the Si5345 via a GPIO and a OBUFDS\_GTE4. , the recovered clocks from mini-FELIX driven to the Si5345 via a GPIO and a OBUFDS\_GTE4.
%local reference clocks

As observed in Figure \ref{fig:ph_orig} and Table \ref{tab:jitter_orig}, the phase noise for the 120 MHz clock is higher than for the 240 MHz clock in the range of frequencies from 10~kHz to 1~MHz, exceeding the manufacturer specifications. However, the tests reveled that the configuration of the Si5345 chip did not cause any relevant effect on the phase noise of the recovered clock indicating that the main jitter contribution is dominated by the CDR of the GBTx.

% Both 240~MHz recovered clocks from FELIX via the GPIO and OBUFDS\_GT4 buffer after the Si5345 jitter cleaner fulfil the manufacturer specifications.

\begin{table}[h!]
\centering
\begin{tabular}{ccc}
% \textbf{}          & \multicolumn{2}{c}{\textbf{Iterations}}                                   \\ \cline{1-3} 
\textbf{Reference}            & \textbf{10~Hz--10~MHz} & \textbf{10~kHz--1~MHz}\\ \hline
\hline
Si5345 120 MHz                         & 3.75~ps & 1.04~ps\\
Si5345 240 MHz                         & 3.57~ps & 0.10~ps\\
\end{tabular}
\caption{Integrated RMS jitter extracted from the measured phase noise plots of the Si5345 driven by the local Si570 oscillator.}
\label{tab:jitter_orig}
\end{table}

%Serenity is intended as the clock-distribution system for at least one system requiring a high-performance(320.624MHz) LHC clock. A further advantage of having the FPGAs mounted on daughter-cards is that the performance of the carrier and interposers can be measured independent of the performance of the FPGA and firmware. A daughter-card was produced to expose the clock-signals:a low-jitter(1.3ps RMS) clock was injected through the ATCA zone-2 connector and the signals measured on the daughter-card using a 6GHz, 20GS/s oscilloscopeconfiguredto acquire 1.3millionclock-cycles with a large acquisition window (20 millionsamples) corresponding to a 1ms  continuous  acquisition,  thereby  scanning  jitter  frequencies  above  1kHz. Across  all  18 available clocks (9 on each interposer site), the channel-to-channel RMS jitter was measured to be 2.8ps,sufficient for Serenity to be considered “an ideal clock-distribution node”.
% \subsection{Firmware structure}

%Downlink
%Uplink
\section{Conclusions and Future Plans}

The new conditions of the HL-LHC implies a complete replacement of the on- and off- detector electronics of the Tile Calorimeter. In addition, a new readout and clocking architecture is needed to fulfill  the requirements of the fully digital trigger for the ATLAS Phase-II Upgrade.

%Clock qualification
The Compact Processing Modules will be the core processing unit of the Tile off-detector electronics at the HL-LHC. The CPMs will interface between the on-detector electronics and the ATLAS TDAQ system, reading out and controlling the on-detector electronics and providing reconstructed cell energy to the ATLAS trigger system in real time.

The first prototypes of the Compact Processing Module for the HL-LHC have been produced and validated. All interfaces have been validated and qualified with BER tests and measuring the optical eye diagrams. Phase noise measurements of the bunch-crossing clock at the CPM present low noise levels which comply with the FPGA manufacturer specifications. The distributed bunch-crossing clock at the test board shows larger phase noise values dominated by the CDR of the GBTx.

% The phase noise measurements of the distributed bunch-crossing clock present high spurs which might need to be mitigated with an improved power filtering circuit on the test board. However, robust communication between the Daughterboard~v5 and the CPM has been demonstrated.

% The phase noise measurements of the bunch-crossing clock recovered at the CPM present low noise levels which fulfil the manufacturer specifications to operate the MGTs. The distributed bunch-crossing clock at the test board presents larger phase noise levels dominated by the CDR of the GBTx.

% The phase noise measurements of the recovered clock in the CPM present low levels which fulfil manufacturer specifications to operate the MGTs, while the distributed bunch-crossing clock extracted by the test board exceeds the phase noise limits which is dominated by the CDR of the GBTx.

% Plans % Demonstrator
Following the ATLAS installation plans for 2025, the production of the CPMs will be divided in two stages: a pre-production of 14 CPMs scheduled for the third quarter of 2021, and a final production of 114 CPM planned by the last quarter of 2023. The complete readout system will be installed in the ATLAS counting rooms in September 2025.

\bibliographystyle{unsrt}
\bibliography{biblio}

\end{document}